%% file: paper.tex
\documentclass[sigconf]{acmart}
\usepackage{array}
\usepackage{pifont}
\usepackage{graphicx}
\usepackage{multirow}
\usepackage[inline]{enumitem}
\usepackage{tikz}
\usepackage{microtype}
\usepackage{xcolor}
\usepackage{ifthen}
\usetikzlibrary{math, calc}

\AtBeginDocument{%
  }

\copyrightyear{2026}
\acmYear{2026}
\setcopyright{cc}
\setcctype{by}
\acmConference[GoodIT '26]{International Conference on Information Technology for Social Good}{September 02--04, 2026}{Pisa, Italy}
\acmBooktitle{International Conference on Information Technology for Social Good (GoodIT '26), September 02--04, 2026, Pisa, Italy}
\acmDOI{10.1145/3794786.3830783}
\acmISBN{979-8-4007-2483-1/2026/09}

\newcommand{\implemented}{\scalebox{1.5}{$\bullet$}}
\newcommand{\discussed}{\scalebox{1.5}{$\circ$}}
\newcommand{\notaddressed}{\scalebox{1.5}{--}}
\newcommand{\codeyes}{\ding{51}}
\newcommand{\codeno}{\ding{55}}

\begin{document}

\title{On the security and privacy of LLMs in Mobility}

\author{Mauro Conti}
\email{mauro.conti@unipd.it}
\affiliation{%
  \city{}
  \institution{University of Padova}
  \country{Italy}
}
\affiliation{%
  \institution{Örebro University}
  \city{Örebro}
  \country{Sweden}
}

\author{Lorenzo Perinello}
\email{lorenzo.perinello@phd.unipd.it}
\affiliation{%
  \institution{University of Padova}
  \country{Italy}
}

\author{Umberto Salviati}
\email{umberto.salviati@phd.unipd.it}
\affiliation{%
  \institution{University of Padova}
  \country{Italy}
}
\affiliation{%
  \institution{Fondazione Bruno Kessler}
  \country{Italy}
}

\renewcommand{\shortauthors}{Conti et al.}

\begin{abstract}
The mobility sector is undergoing a paradigm shift driven by advances in Generative Artificial Intelligence. With a global market valued at approximately 2.9 trillion dollars annually, considering only cars, the integration of these technologies has the potential to impact more than 1.5 billion vehicles worldwide. As Large Language Models (LLMs) are increasingly adopted in mobility, concerns about cybersecurity, privacy, and reliability emerge. Accordingly, this paper surveys current applications and assesses these challenges.
Since the European AI Act classifies transportation AI as high risk, we derive nine technical classes from its requirements to assess current research and future deployments.
Our findings show that research mainly studies GPT and Llama models (over 50\% of reviewed works) and traffic applications while largely neglecting security, privacy, and reliability. This gap extends to AI Act compliance: among 35 reviewed works, only one includes a partial vulnerability assessment and one a partial risk management system. We identify a clear gap between strong optimization performance and regulatory adherence, suggesting compliance is limited less by technology than by a focus on static performance over lifecycle safety, and underscoring an urgent need for security-by-design in safety-critical intelligent transportation systems.
\end{abstract}



\maketitle

\newpage
\section{Introduction}
Mobility is a ubiquitous concept related to a multitude of social and economic aspects: thereby, it represents an umbrella term encompassing multiple stakeholders and sectors and, today more than ever, raises questions and opportunities~\cite{adey2017mobility}.

Focusing on urban mobility, defined as the movement of individuals within urban areas, which currently host nearly 5 billion people worldwide, issues and challenges concerning the efficiency and effectiveness of mobility management are becoming increasingly significant ~\cite{UNUrbanization2025}. The importance of urban mobility is also reflected in the growing body of scientific literature on the topic, which explores the broader sector, investigates its limitations and challenges, and focuses on its various dimensions ~\cite{MISKOLCZI2021103029}. One of these dimensions concerns Intelligent Transport Systems (ITS), also referred to as smart mobility: a framework that integrates information and communication technologies, and applies them to the transport sector ~\cite{perallos2015intelligent}. However, such integration has also introduced new security and privacy challenges, including cyber threats, data protection issues, and risks associated with the reliability and trustworthiness of these systems. Consequently, security has become a fundamental aspect in the design, deployment, and management of modern mobility systems~\cite{Abdo202410292878}.

At the same time, artificial intelligence (AI) has emerged as a novel and important technology within the mobility sector and has been widely exploited for optimization, simulation, and prediction tasks within ITS. More recently, generative AI technologies, such as LLMs, have emerged as key enabling technologies for these tasks ~\cite{Mahmud202510851302}. Furthermore, autonomous AI systems (Agentic AI) have also been integrated into the mobility domain to boost the effectiveness of simulations, management processes, and other mobility-related tasks ~\cite{Dokuchaev10297009}. While researchers and industry drivers focus almost exclusively on enhancing the performance ~\cite{pati2025agentic}, accuracy, and operational capabilities of Generative Artificial Intelligence in transport, they systematically overlook the foundational principle of security-by-design ~\cite{tallam2025engineering}. As a consequence, the rapid deployment of LLMs and autonomous agents effectively creates an entirely new threat surface that can be readily exploited by malicious actors ~\cite{dong2024attacks}.

Recognizing the severity of these emerging vulnerabilities, regulatory bodies ~\cite{NIST_AIRMF, EU_AI_Act_2024} have increasingly expressed deep concern about the rapid deployment of generative models in critical sectors. This institutional awareness of the threat is manifested in the enforcement of the European AI Act ~\cite{EU_AI_Act_2024}, which will become fully applicable by August 2026, with specific compliance deadlines for embedded high-risk mobility systems extending to August 2027. Under this framework, regulatory entities explicitly acknowledge that AI applications integrated into critical infrastructure (i.e., urban transportation and traffic management systems) pose substantial systemic risks, frequently classifying them as ``high-risk'' systems. The AI Act mandates strict compliance regarding technical robustness, cybersecurity, and continuous human oversight. Consequently, understanding the intersection of generative AI vulnerabilities and regulatory mandates is no longer just a technical precaution but a legal requirement.

To the best of our knowledge, no study has yet systematically investigated the current trends and trajectories of LLMs within the mobility context through a security-focused lens. This paper aims to bridge this gap. We provide a comprehensive security analysis of LLMs deployment in mobility systems, identifying current gaps, mapping the expanding threat landscape, and aligning these technical risks with the compliance requirements set by the AI Act.

Therefore, the contributions of this paper are to provide a targeted survey of the topic and to investigate specific security, privacy and reliability aspects of the works taken into account. Additionally, the work presents a translation of the EU AI Act provisions into 9 technical evaluation criteria for AI transportation systems, which are then systematically assessed against the investigated works.

This work is organized as follows: Section \ref{sec:rel_works} provides an overview of related work, while Section \ref{sec:survey} describes the survey, the adopted methodology, and the findings. Section \ref{sec:aiAct} describes our AI Act related contribution. Finally, \ref{sec:conclusion} outlines the conclusion and directions for future works.

\section{Related Work}
\label{sec:rel_works}
This section gives a brief overview of security in AI and summarizes recent surveys on AI in transportation.

\paragraph{Security on AI}
A June 2026 Google Scholar search for \textit{(``large language model'' OR ``llm'') AND (``security'' OR ``vulnerability'' OR ``attack'')} returned over 32,000 results. Since 2022 (before the rollout of ChatGPT), more than 420 LLM security papers have appeared in top venues (NDSS, IEEE S\&P, ACM CCS, USENIX Security), indicating rapid progress in LLM security research. AI security has traditionally been studied as Adversarial Machine Learning (AML), which examines the security, vulnerabilities, and robustness of AI systems across their lifecycle, including both AI-level and system-level exploits ~\cite{HTPT, light}. 
Focusing specifically on LLMs, since it is the goal of our work, introduces risks unique to generative models. The main threats are jailbreaking ~\cite{yi2024jailbreak} and prompt injection ~\cite{kumar2024strengthening}. Jailbreaking bypasses alignment guardrails to elicit prohibited content, while prompt injection uses text-based adversarial examples to override system constraints. Indirect prompt injection is especially dangerous, as malicious payloads are silently ingested from untrusted external sources such as parsed documents or web pages ~\cite{Misleading}. 

\paragraph{AI in transportation} Recent years have witnessed growing interest in the integration of Generative AI and LLMs into mobility systems, resulting in numerous studies investigating their applications. While a few works examine broader contexts, such as the role of Generative AI in smart cities~\cite{Pradhan11125750} and digital twins~\cite{Duran202411101058}, which may also encompass mobility-related applications and considerations, most studies focus specifically on transportation-related domains. For example, Mahmud et al.~\cite{Mahmud202510851302} provide a comprehensive review of LLM integration in ITS, covering applications such as traffic prediction, traffic signal optimization, autonomous driving, and vehicle and pedestrian detection. The survey highlights the potential of LLMs to improve transportation efficiency and decision-making while also discussing technical and deployment challenges. Other works survey the application of LLMs in ITS: Kaur et al.~\cite{KAUR2026100308} synthesize the LLM models, methods, and datasets used in the literature, while Hassija et al.~\cite{HASSIJA2026100996} present a broader, application-oriented survey of LLMs in ITS, discussing their adoption in traffic management, autonomous driving, predictive maintenance, vehicular communications, and the challenges associated with integrating multimodal data. Further studies focus on specific aspects of LLMs in mobility: Jin et al. ~ ~\cite{Jin20022026} examine LLMs from a transportation planning perspective, covering transportation analysis, modeling, and prediction, while Zhang and Dingkai~\cite{ZhangDingkai2024} focus specifically on transportation management and safety.

Overall, existing papers provide valuable insights into the usage of LLMs for mobility applications, however, none of the previously cited works systematically addresses security (or related fields) considerations or analyzes the security challenges arising from their adoption. As a result, despite its importance, security remains a largely overlooked research area within the current body of literature.

\section{Survey}
\label{sec:survey}
This section presents a preliminary survey of the literature on the application of LLMs in urban mobility, with a twofold goal: to analyze the models used and the tasks addressed, and to examine security, privacy, reliability and reproducibility aspects of these systems. Finally the findings, which are discussed in Section \ref{sec:findings}, provide an overview of the current landscape and identify research gaps that motivate further investigation into secure and trustworthy generative AI solutions for urban mobility.

\subsection{Methodology}
We conducted a targeted literature search in Web of Science using a boolean query that combined mobility terms with LLM-related terms. Guided by Google Trends data for the first half of 2026, we compared major generative AI platforms: ChatGPT, Google Gemini, Claude AI, Microsoft Copilot, Grok AI, Llama, Qwen, Microsoft Phi, Mistral AI, and DeepSeek. ChatGPT \cite{achiam2023gpt} showed the highest relative search interest (average interest index \footnote{For the average interest definition, see \url{trends.google.com/trends}}: 92), while Llama and Gemini ranked much lower, 8 and 6, respectively, and the others showed negligible interest. Although Google Trends does not reflect actual usage, we used relative search interest as proxy data for public attention and visibility, motivating our focus on GPT-based architectures. Our mobility query included: urban mobility, intelligent transport system(s), smart mobility, traffic management, spatio-temporal, urban computing, urban sensing, smart city, and urban prediction, combined with OR. This block was then linked with AND to at least one generative AI term: large language model(s), generative artificial intelligence, agentic AI, autonomous agent(s), LLM, or GPT. The search returned 125 publications. After removing duplicates, review papers (see Section~\ref{sec:rel_works}), and out-of-scope studies, we obtained 35 papers for full-text analysis.

\subsection{Key elements inspected}
Table~\ref{table:survey} summarizes the studies included in the review, reporting the functionality of the models utilized, classified according to the taxonomy provided in ~\cite{NIE2025100003}: Information Processor (A),  Knowledge Encoders (B), component generators (C) and decision facilitators (D). A includes papers that use AI as context encoder, data analyzer or a multimodal fuser; B includes knowledge extractor and representation embedder; C includes simulator, evaluator and interpreter, or data synthesizer; finally D includes Decision maker or guider and spatio-temporal predictors. A more detailed description of these categories is provided in~\cite{NIE2025100003}.
Moreover, we reported the task addressed, and the extent to which security, privacy, and reliability aspects are considered. These aspects are categorized as follows:

\begin{itemize}
    \item \textbf{discussed} (\discussed): the topic is mentioned or discussed in the paper, either briefly or in detail, but no concrete implementation is provided;
    
    \item \textbf{implemented} (\implemented): the paper not only discusses the topic but also describes and incorporates specific mechanisms, methods, or procedures addressing it;
    
    \item \textbf{not addressed} (\notaddressed): the topic is neither discussed nor implemented within the study.
\end{itemize}

\input{main_table}

\subsection{Findings}
\label{sec:findings}
This section provides an overview of the findings derived from the preliminary survey, highlighting the most commonly adopted models and application tasks, as well as the extent to which security, privacy, reliability, and reproducibility aspects are considered in current research.

\subsubsection{Models}
\begin{figure}[h]
    \centering
    \includegraphics[width=0.8\linewidth]{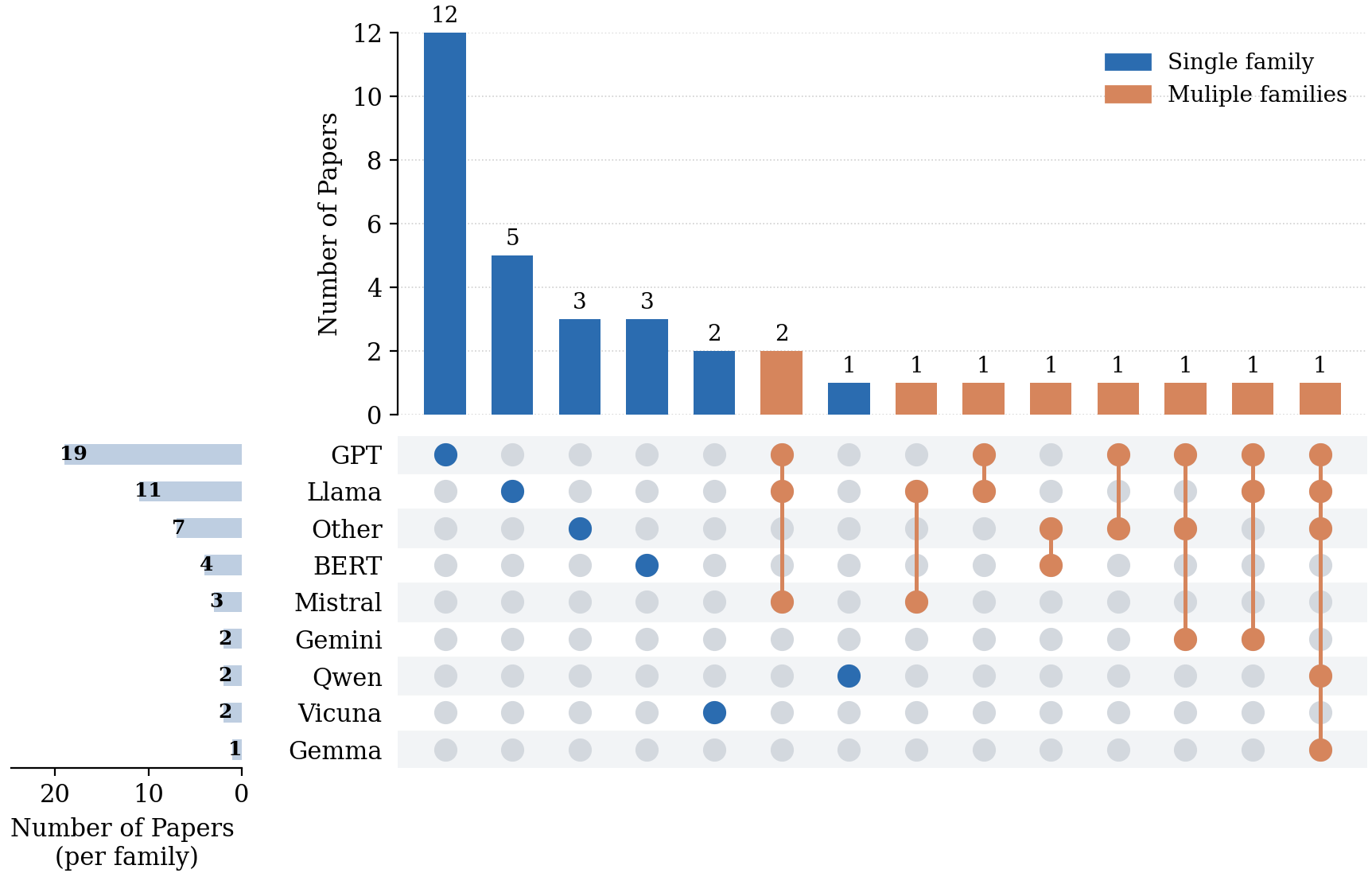}
    \caption{Distribution of Model Families Across the Reviewed Papers}
    \Description{Distribution of Model Families Across the Reviewed Papers}
\label{fig:model_families}
\end{figure}

Figure \ref{fig:model_families} summarizes the distribution of model families employed across the reviewed studies: the GPT family appears in 19 papers, accounting for more than half of the reviewed papers; the Llama family is the second-most-frequently used category, appearing in 11 studies. In contrast, language models such as BERT are less common, appearing in only four papers. Other large language models are represented only marginally in the literature:  Mistral family appears in three papers, and Gemini, Qwen, Gemma, and Vicuna occur in two, and Vicuna appears in one. Other models, which represent the remaining seven paper, include other LLMs such as DeepSeek or Chronos.

\subsubsection{Functionality}
\label{subsec:func}
Decision facilitators emerged as the dominant functionality category, appearing in 25 of the 35 reviewed studies (71.4\%). Then, Information Processor was the second most common, occurring in 18 studies (51.4\%), while Knowledge Encoders appeared in 11 studies (31.4\%). In contrast, we identify Functionality C in only one publication. Notably, functionalities frequently co-occurred, particularly A and D, which appeared together in 10 studies. 

\subsubsection{Security, Privacy, Reliability}
\label{sec:security}
Regarding the acknowledgment of security aspects in the scrutinized papers, only two studies implemented security measures at least partially within their proposed methodologies, while an additional four papers merely discussed security-related concerns without providing concrete implementation details. 
Among the measures implemented, ~\cite{ZHANG202495} maintains data security and integrity, and ~\cite{Kalantari2024} prevents the use of invalid or harmful prompts ~\cite{Muriuki2024AdvancedIT}. Other works, ~\cite{Muriuki2024AdvancedIT, s25216698, Kalyuzhnaya2025, ChenYong2025, HaoZhang11045306}, only discuss security-related problems without any explicit implementation.

With respect to privacy, only ~\cite{FAN2026131620} explicitly addresses privacy preservation by explicitly taking into account data privacy in the selection of the datasets used;  six additional studies (\cite{ZHANG202495, Muriuki2024AdvancedIT, s25216698, ChenYong2025, Bhandari2024, Masri2024LeveragingLL, Beneduce10981725}) discuss only privacy at high level: ~\cite{ZHANG202495, ChenYong2025} state privacy without giving any implementation details, others  (~\cite{Kengne2025, s25216698, Calderon2025s25185688}) declare privacy as future work trajectory and only ~\cite{Masri2024LeveragingLL} specifies clearly privacy as a limitation of the work.

Regarding reliability, six studies incorporate mechanisms to improve the robustness and dependability of their proposed systems. In \cite{ZHANG202495}, the LLM-based system is equipped with mechanisms to ensure data authenticity, and incorporate a human-in-the-loop intervention protocol by design. Furthermore, \cite{Syum202511194151} improves reliability by implementing output validation procedures. In \cite{ChenYile2025} and \cite{LiZhonghang2024}, reliability is assessed by evaluating the proposed frameworks with different LLM backbones, demonstrating consistent performance across models. Furthermore, \cite{Kalantari2024} explicitly states the evaluation of the robustness as an objective of the paper. Finally, \cite{LouShangyu2025} enhances robustness by employing output comparison, consistency checks, and selective re-extraction mechanisms to mitigate bias and improve the reliability of generated predictions. All the other papers that discuss reliability address it as a limitation or future work.

\subsubsection{Code Availability} Regarding the open-source aspect, which is crucial for reproducibility of results and for conducting security and related research, among the investigated papers, 13 out of 35 made the code available through a repository.

\section{AI Act in Transportation Reserach}
\label{sec:aiAct}
This section contextualizes the current trajectories of LLMs in urban mobility within the newly established European regulatory framework \cite{EU_AI_Act_2024}. By mapping the technical vulnerabilities discovered in literature to explicit legal obligations, we evaluate the systemic readiness of the intelligent transportation research domain.

\subsection{Risk Classification in Transportation Systems}

The AI Act introduces a strict risk-based paradigm \cite{eu_ai_act_framework_web} that categorizes artificial intelligence systems into four distinct levels based on their potential to cause harm, establish systemic vulnerabilities, or infringe upon fundamental rights:

\begin{enumerate*}[label=(\roman*)]
    \item \textit{Unacceptable Risk:} AI systems that pose a clear threat to the safety, livelihoods, and rights of individuals are strictly prohibited. This category includes practices such as social scoring, harmful AI-based manipulation, and real-time remote biometric identification in publicly accessible spaces.
    \item \textit{High Risk:} Use cases that can pose serious risks to human health, safety, or fundamental rights are classified under this tier. These systems are subject to strict legal and technical obligations before they can be deployed on the market.
    \item \textit{Limited (Transparency) Risk:} This category applies to AI systems with a specific need for disclosure, such as chatbots \cite{openai2026chatgpt, google2026gemini, anthropic2026claude, deepseek2026, perplexity2026, microsoft2026copilot, meta2026ai} or generative AI tools \cite{midjourney2026, openai2026dalle3}. Providers must ensure that humans are explicitly informed they are interacting with an automated system and that AI-generated content or deep fakes are clearly identifiable.
    \item \textit{Minimal or No Risk:} The vast majority of AI applications currently utilized within the EU (e.g., spam filters or AI-enabled video games) fall into this category and face no regulatory restrictions under the Act.
\end{enumerate*}

Under this risk-based paradigm, AI safety components in critical transportation infrastructure, such as road traffic management and transport operations, are classified as high-risk under Article 6(2)\cite{eu_ai_act_article6} and Annex III \cite{EU_AI_Act_2024, eu_ai_act_annex3}, as their failure could endanger public health and safety \cite{EU_AI_Act_2024}. As discussed in our literature survey (Section \ref{sec:survey}), recent research increasingly uses LLMs and autonomous agentic architectures for core mobility tasks such as real-time traffic management, traffic light optimization, and vehicle dispatching.

\subsection{Compliance Framework Evaluation Criteria}
While these top-down institutional tools offer a generalized baseline, mobility systems present highly interconnected threat surfaces bridging General Purpose AI (GPAI) \cite{europeancommission2026codeofpractice} and embedded safety components. To address this domain-specific complexity, we translate the high-level legal articles of the AI Act into actionable development and evaluation requirements tailored for the transportation sector, as structured in Table~\ref{table:eu_ai_act}.

\begin{table*}[t]
\centering
\footnotesize
\setlength{\tabcolsep}{4pt}
\renewcommand{\arraystretch}{1.15}

\begin{tabular}{%
    p{3.6cm}
    p{7.2cm}
    p{3.8cm}
    p{2.2cm}
}
\toprule
\textbf{Category} &
\textbf{Requirement} &
\textbf{Article Reference} &
\textbf{Mean $\pm$ SD} \\
\midrule

\multirow{3}{*}{\parbox{3.4cm}{\raggedright Vulnerability Evaluation\\ (GPAI w/ Systemic Risk)}}
    & Adversarial testing (assessment)          & Art.\ 55(1)(a)            & $1.11 \pm 0.53$ \\
    & Risk mitigation measures (implementation) & Art.\ 55(1)(b),(c)        & $1.06 \pm 0.34$ \\
    & Cybersecurity protection                  & Art.\ 55(1)(d)            & $1.17 \pm 0.62$ \\
\midrule

\multirow{4}{*}{\parbox{3.4cm}{\raggedright Risk Management System\\ (High-Risk)}}
    & Lifecycle risk identification \& analysis & Art.\ 9(1), Art.\ 9(2)(a), Art.\ 9(2)(c) & - \\
    & Misuse scenario evaluation                & Art.\ 9(2)(b), Art.\ 9(9)                & - \\
    & Residual risk acceptability               & Art.\ 9(5)                               & - \\
    & Real-world/pipeline attack testing        & Art.\ 9(6), Art.\ 9(7), Art.\ 9(8), Art.\ 60 & $1.09 \pm 0.37$ \\
\midrule

\multirow{5}{*}{\parbox{3.4cm}{\raggedright Quality Management System\\ (High-Risk)}}
    & Design controls, test plans               & Art.\ 17(1)(b--d)         & - \\
    & Data governance                           & Art.\ 17(1)(f), Art.\ 10  & - \\
    & Logging by design                         & Art.\ 12, Art.\ 19(1)     & - \\
    & Post-market monitoring                    & Art.\ 17(1)(h), Art.\ 72  & - \\
    & Incident reporting                        & Art.\ 17(1)(i), Art.\ 73(1) & - \\
\midrule

\multirow{5}{*}{\parbox{3.4cm}{\raggedright Transparency \& Copyright}}
    & Interaction disclosure (chatbot)          & Art.\ 50(1)               & - \\
    & AI-generated content watermarking         & Art.\ 50(2)               & $1.03 \pm 0.17$ \\
    & Deep fake labelling                       & Art.\ 50(4)               & - \\
    & Training data summary (GPAI)              & Art.\ 53(1)(d)            & $1.00^{\dagger}$ \\
    & Copyright compliance policy (GPAI)        & Art.\ 53(1)(c)            & - \\
\midrule

\multirow{2}{*}{\parbox{3.4cm}{\raggedright Technical Documentation}}
    & Compliance for Regulators                 & Art.\ 53(1)(a), Annex XI  & - \\
    & Transparency for Downstream Users         & Art.\ 53(1)(b), Annex XII & - \\
\midrule

\multirow{2}{*}{\parbox{3.4cm}{\raggedright Human Oversight\\ (High-Risk)}}
    & Override/interrupt mechanism               & Art.\ 14(3), Art.\ 14(4)(d--e)           & - \\
    & Automation bias \& cognitive safeguards   & Art.\ 14(1--3); Art.\ 14(4)(a--c); Art.\ 14(5) & - \\
\midrule

\multirow{2}{*}{\parbox{3.4cm}{\raggedright Accuracy \& Robustness\\ (High-Risk)}}
    & Defined accuracy metrics                  & Art.\ 15(1), 15(2), 15(3)                 & \textbf{3.03 $\pm$ 0.79} \\
    & Robustness and adversarial resilience     & Art.\ 15(1), 15(4), Art.\ 15(5)           & $1.57 \pm 1.12$ \\
\midrule

\multirow{2}{*}{\parbox{3.4cm}{\raggedright Fundamental Rights \& Deployment}}
    & FRIA conducted before deployment          & Art.\ 27                  & - \\
    & Informing workers of AI deployment        & Art.\ 26                  & - \\
\midrule

\multirow{3}{*}{\parbox{3.4cm}{\raggedright Compliance \& Administrative\\ (High-Risk)}}
    & Conformity assessment procedure           & Art.\ 43                  & - \\
    & Registration in the EU database           & Art.\ 49                  & - \\
    & EU Declaration of Conformity \& CE marking & Art.\ 47, Art.\ 48       & - \\

\bottomrule
\end{tabular}
\begin{flushleft}
\scriptsize $^{\dagger}$ indicates that the standard deviation value is equal to zero ($\text{std} = 0.00$).
\end{flushleft}
\caption{EU AI Act Compliance Requirements with paper evaluation means and standard deviations.}
\label{table:eu_ai_act}
\end{table*}

To systematically assess the selected state-of-the-art mobility literature against the technical mandates of Regulation (EU) 2024/1689 (the AI Act), this section examines each requirement presented in Table~\ref{table:eu_ai_act} item by item. Specifically, we detail the underlying operational objectives for each criterion, defining the baseline compliance scope used to evaluate the reviewed works:
\noindent \paragraph{\textbf{Vulnerability evaluation}}
\begin{itemize}
    \item \textit{Adversarial Testing:} The execution and documentation of standardized adversarial testing protocols to proactively identify vulnerabilities, safety weaknesses, and potential systemic risks within the AI model.
    \item \textit{Risk Mitigation Measures:} The strategic assessment of systemic risks at the Union level, coupled with the implementation of mitigation measures and the procedural management of serious incidents and corrective actions.
    \item \textit{Cybersecurity Protection:} The assurance of an adequate level of security for both the AI model and its underlying physical infrastructure to prevent unauthorized access, manipulation, or compromise.
\end{itemize}

\noindent \paragraph{\textbf{Risk management system}}
\begin{itemize}
    \item \textit{Lifecycle Risk Identification \& Analysis:} Continuous, iterative management of risks across the AI system's entire lifecycle, including initial analysis and post-market monitoring feedback loops.
    \item \textit{Misuse Scenario Evaluation:} Evaluation of risks arising from reasonably foreseeable misuse, including impacts on fundamental rights and specific safeguards for vulnerable groups or minors.
    \item \textit{Residual Risk Acceptability:} Formal application of the mitigation hierarchy (design, controls, information and training) to reduce risks, followed by a documented judgment of residual risk acceptability.
    \item \textit{Real-World / Pipeline Attack Testing:} Technical validation through predefined metrics or thresholds and, if applicable, Art. 60-compliant \cite{EU_AI_Act_2024} real-world testing plans, informed consent, and regulatory oversight.
\end{itemize}

\noindent \paragraph{\textbf{Quality management system}}
\begin{itemize}
    \item \textit{Design Controls \& Test Plans:} The implementation of systematic actions, technical procedures, and validation processes used for the design, development, and verification of the high-risk AI system.
    \item \textit{Data Governance:} Management practices for data acquisition, collection, labelling, storage, and bias mitigation, ensuring data sets are representative and of high quality.
    \item \textit{Logging by Design:} Technical capabilities allowing for the automatic, continuous recording of events throughout the AI system's lifetime to ensure traceability and operational accountability.
    \item \textit{Post-Market Monitoring:} A systematic process to actively collect, document, and analyze data on the system’s performance after it has been placed on the market.
    \item \textit{Incident Reporting:} Procedures for the identification, investigation, and timely notification of serious incidents to national competent authorities.
\end{itemize}

\noindent \paragraph{\textbf{Transparency \& copyright}}
\begin{itemize}
    \item \textit{Interaction Disclosure:} The implementation of clear, distinguishable mechanisms to inform natural persons that they are interacting with an AI system rather than a human, unless it is obvious from the context.
    \item \textit{AI-Generated Content Watermarking:} The technical implementation of machine-readable, robust, and interoperable markings on synthetic audio, image, video, or text outputs to ensure they are detectable as artificially generated.
    \item \textit{Deep Fake Labelling:} Obligations for deployers to disclose when content constitutes a deep fake (artificially generated or manipulated imagery, audio, or video) or to disclose the use of AI-generated text in public interest publications.
    \item \textit{Training Data Summary:} The creation and public disclosure of a sufficiently detailed summary of the training content used for General-Purpose AI models, following the template provided by the AI Office.
    \item \textit{Copyright Compliance Policy:} The establishment and execution of an internal policy to respect Union copyright law, specifically identifying and adhering to the ``reservation of rights'' (opt-outs) expressed by rightsholders.
\end{itemize}

\noindent \paragraph{\textbf{Technical documentation}}
\begin{itemize}
    \item \textit{Compliance for Regulators:} The comprehensive collection of technical documentation regarding the model’s development, training, testing processes, and evaluation results, maintained for the AI Office and national competent authorities.
    \item \textit{Transparency for Downstream Users:} The documentation provided to downstream AI system providers, detailing the model’s capabilities and limitations to ensure safe integration and compliance with regulatory obligations.
\end{itemize}

\noindent \paragraph{\textbf{Human oversight}}
\begin{itemize}
    \item \textit{Override / Interrupt Mechanism:} The implementation of design features and procedures that allow natural persons to effectively monitor the system, intervene in its operation, and safely halt or override its output or functioning when necessary.
    \item \textit{Automation Bias \& Cognitive Safeguards:} The implementation of tools, training, and processes (including dual-verification where required) that enable natural persons to understand system limitations, avoid over-reliance on AI outputs, interpret results correctly, and perform independent validation.
\end{itemize}

\noindent \paragraph{\textbf{Accuracy \& robustness}}
\begin{itemize}
    \item \textit{Defined Accuracy Metrics:} The systematic process for defining, measuring, and documenting the accuracy performance of the AI system, including the selection of appropriate metrics and their communication in user instructions.
    \item \textit{Robustness and Adversarial Resilience:} The implementation of technical and organizational measures to ensure the system performs consistently under stress, handles errors gracefully, and remains secure against malicious attempts to manipulate performance or exploit vulnerabilities (e.g., data poisoning, evasion).
\end{itemize}

\noindent \paragraph{\textbf{Fundamental rights}}
\begin{itemize}
    \item \textit{Fundamental Rights Impact Assessment (FRIA) Conducted Before Deployment:} The process of identifying, assessing, and documenting the potential impact of a high-risk AI system on fundamental rights, including the mitigation of identified risks and the implementation of human oversight measures.
    \item \textit{Informing Workers of AI Deployment:} The transparency and communication protocol established to inform workers and their representatives about the deployment of high-risk AI systems in the workplace, ensuring compliance with Article 26(7).
\end{itemize}

\noindent \paragraph{\textbf{Compliance \& administrative}}
\begin{itemize}
    \item \textit{Conformity Assessment Procedure:} The process of verifying that the high-risk AI system complies with the requirements set out in the AI Act, utilizing either internal control or a third-party notified body assessment as mandated by Article 43.
    \item \textit{Registration in the EU Database:} The mandatory registration of the high-risk AI system (and in some cases, the deployer) in the EU database pursuant to Article 49 before the system is placed on the market or put into service.
    \item \textit{EU Declaration of Conformity \& CE Marking:} The creation of the written EU Declaration of Conformity (Art. 47) and the affixing of the CE marking (Art. 48) to indicate the AI system’s adherence to all relevant Union harmonization legislation.
\end{itemize}

\subsection{Compliance Assessment of the Reviewed Work}
While the previous section defines the regulatory framework of the EU AI Act for high-risk AI systems, analyzing the exploratory literature requires a more focused and quantifiable methodological scope. To establish a rigorous empirical baseline, the corpus of $35$ selected papers has been systematically evaluated across nine distinct technical criteria derived from the regulatory mandates. This cross-sectional analysis yields a total dataset of $280$ individual evaluation points ($35 \text{ papers} \times 8 \text{ criteria}$), providing a comprehensive mapping of the current research landscape against the European framework. 

\subsubsection{Scope Selection}
Assessing academic research under the EU AI Act requires separating technical design from deployment. Since this study analyzes algorithmic architectures rather than commercial systems, applying the full regulatory spectrum is inappropriate. 

Our scope isolates eight requirements within the researcher's sphere of influence, mapping to the \textit{Security and Compliance by Design} paradigm across four framework categories:
\begin{enumerate}[label=(\roman*)]
    \item \textit{Vulnerability Evaluation (GPAI with Systemic Risk)}: \textit{Adversarial testing}, \textit{Risk mitigation}, and \textit{Cybersecurity protection}, addressing model flaws and execution security.
    \item \textit{Risk Management System (High-Risk)}: \textit{Real-world/pipeline attack testing} to evaluate pipeline resilience before industrialization.
    \item \textit{Transparency \& Copyright}: \textit{AI content watermarking} and \textit{Training data summary}, embedding traceability into generation routines.
    \item \textit{Accuracy \& Robustness (High-Risk)}: \textit{Defined accuracy metrics} and \textit{Adversarial resilience}, forming the core of empirical benchmark testing.
\end{enumerate}

Conversely, downstream operational and administrative duties, such as Quality Management Systems (QMS), Technical Documentation with compliance documentation, Human Oversight, Fundamental Rights and Deployment, and Compliance and Administrative categories are excluded. These represent institutional governance belonging strictly to commercial deployment.

\subsubsection{Evaluation Metric and Likert Scale}
To evaluate the depth to which each reviewed work addresses these fundamental legal mandates, we apply a standardized 5-point Likert scale. This metric measures the progression from total regulatory omission to full pipeline integration for each requirement:
\begin{enumerate}
    \item[\textbf{1.}] \textbf{Not Mentioned:} There is no evidence, acknowledgment, or discussion of the requirement or its underlying safety implications.
    \item[\textbf{2.}] \textbf{Identified:} The requirement or threat surface is explicitly acknowledged as a theoretical necessity, but no actionable technical methodology or concrete test cases are provided.
    \item[\textbf{3.}] \textbf{Assessed:} The requirement is actively evaluated or tested as a static, one-time baseline activity (e.g., standard static testing), but lacks a formalized protocol or structured lifecycle integration.
    \item[\textbf{4.}] \textbf{Planned / Implemented:} Structured technical mechanisms, defensive policies, or compliant testing plans are formally implemented and fully documented within the proposed architecture.
    \item[\textbf{5.}] \textbf{Fully Assessed \& Integrated:} Advanced, automated compliance workflows are seamlessly integrated into the development framework.
\end{enumerate}

\subsubsection{Empirical Evaluation}

The empirical mapping of the reviewed papers reveals a pronounced empirical divergence between high functional performance and near-absent real security compliance (see Table \ref{table:eu_ai_act}). Academic research focuses almost exclusively on the applications themselves, systematically omitting the technical and procedural safeguards mandated by Regulation (EU) 2024/1689.

\begin{itemize}
    \item \textbf{Accuracy \& Robustness [$\mu = 2.30$, $\sigma = 1.21$, $N = 70$]:} This category features the highest macro average but remains heavily polarized. While researchers meticulously track standard performance parameters ($\mu = 3.03$, $\sigma = 0.79$), actual resilience against cyberattacks drops sharply ($\mu = 1.57$, $\sigma = 1.12$). This observation highlights a clear tendency to conflate statistical precision with structural safety. However, the high variance points to rare, virtuous exceptions ($\max = 5$) capable of embedding effective defensive barriers, proving that the current bottleneck is a matter of research priorities rather than technical feasibility.

    \item \textbf{Vulnerability Evaluation [$\mu = 1.11$, $\sigma = 0.51$, $N = 105$]:} A critical blind spot for the domain. Near-zero scores in cybersecurity protection ($\mu = 1.17$, $\sigma = 0.62$) and vulnerability testing ($\mu = 1.11$, $\sigma = 0.53$) confirm a widespread lack of proactive controls against malicious exploitation. Only isolated exceptions ($\max = 5$) demonstrate that subjecting models to stress tests and attack simulations prior to publication is entirely achievable.

    \item \textbf{Risk Management System [$\mu = 1.09$, $\sigma = 0.37$, $N = 35$]:} Risk governance is virtually non-existent. Real-world pipeline attack testing and risk mitigation measures share an identical low rating ($\mu = 1.09$, $\sigma = 0.37$ and $\mu = 1.06$, $\sigma = 0.34$, respectively). Even the top-scoring studies ($\max = 5$) limit security to basic technical bug-tracking, completely omitting organizational governance, contingency fallback plans, and the regulatory notification channels required by law.

    \item \textbf{Transparency \& Copyright [$\mu = 1.01$, $\sigma = 0.12$, $N = 70$]:} Total stagnation across the entire sector. The inclusion of digital watermarking in synthetic content is nominal ($\mu = 1.03$, $\sigma = 0.17$). The most critical metric is training data transparency, which registers absolute zero variance ($\mu = 1.00$, $\sigma = 0.00$, $\max = 5$): none of the evaluated papers provide or plan copyright-compliant dataset documentation, marking a collective failure regarding transparency obligations.
\end{itemize}

\section{Conclusion and Future Directions}
\label{sec:conclusion}
This paper investigated the integration of Large Language Models within the mobility domain by analyzing 35 studies through a security lens and evaluating their compliance readiness against the European AI Act. 
Findings indicate a technological landscape dominated by the GPT and Llama model families, which together account for more than half of the surveyed papers. Such concentration not only reflects the popularity of these models in current research but also raises concerns that vulnerabilities affecting them could propagate across a large portion of mobility applications. Furthermore, only five papers discuss security-related aspects, and only two incorporate security mechanisms into their proposed solutions. This trend suggests that security remains largely overlooked by the research community, despite its critical importance for mobility applications. Consequently, the real-world deployment of many proposed approaches could be premature without a comprehensive security assessment and validation. This also applies to security as well: only one paper implements countermeasures, and seven discuss this aspect mainly as future work, limitations, or without providing any real details. The very limited attention devoted to privacy suggests that the design of LLM-based mobility solutions often lacks concrete privacy-preserving principles and mechanisms. Although reliability is the most widely implemented element in the works (six papers out of 35 implement reliability-related mechanisms), hallucination prevention, output validation, human-in-the-loop, and other mitigation methods are clearly underinvestigated and underimplemented in a critical sector such as mobility. Overall, the survey highlights systematic vulnerabilities in the mobility domain, providing fertile ground for potential attacks and threats. These findings align with the compliance categories of the EU Regulation 2024/1689 discussed in Section \ref{sec:aiAct}. The Act strictly regulates AI-based transportation systems, classifying the mobility sector as a high-risk AI application area. Quantitative analysis shows a statistically significant gap between high functional performance and regulatory alignment, indicating that current literature prioritizes capabilities over required technical and procedural safeguards. Although isolated peak scores ($\max = 5$) prove that compliance is technically feasible, they are rare: this suggests the main barrier is not technical capability, but the research community’s treatment of safety and security beyond deployability. Proactive vulnerability testing, lifecycle risk governance, and copyright-compliant training data transparency are largely absent from most worsk.

With this work, which underlines a clear gap between the adoption of LLMs in mobility and the treatment of security and privacy. First, developing interdisciplinary frameworks at the intersection of legal compliance and transportation AI is critical to translate regulatory mandates into quantifiable operational constraints. Second, establishing robust deployability paradigms will ensure that LLMs can handle real-time, large-scale mobility data safely at the edge. Third, future work must embed advanced cybersecurity mechanisms directly into Intelligent Transportation Systems to secure the entire AI lifecycle against data manipulation. Finally, achieving full compliance with overarching frameworks like the AI Act requires strict adherence to copyright laws for transit datasets, alongside rigorous, real-world stress testing. This demands continuous adversarial risk assessment measures to guarantee systemic resilience against prompt injections, data poisoning, and cyber threats in physical deployment.

\footnotesize{
\begin{acks}
This work is funded by the MUR
(Italian Ministry of Universities and Research)  under the
MOST—Sustainable Mobility National Research Center initiative within
the European Union Next-Generation EU (PIANO NAZIONALE DI RIPRESA E
RESILIENZA (PNRR)—MISSIONE 4 COMPONENTE 2, INVESTIMENTO 1.4—D.D. 1033
17 June 2022, CN00000023), as well as from the Department of
Mathematics of the University of Padua.
\end{acks}

\bibliographystyle{acm}
\bibliography{refs}

\end{document}

%% file: main_table.tex
\begin{table*}[t]
\centering
\footnotesize
\setlength{\tabcolsep}{3pt}
\renewcommand{\arraystretch}{1.15}

\begin{tabular}{%
    >{\raggedright\arraybackslash}p{1.0cm}
    >{\raggedright\arraybackslash}p{2.8cm}|
    >{\raggedright\arraybackslash}p{0.7cm}
    >{\raggedright\arraybackslash}p{4.0cm}|
    >{\centering\arraybackslash}p{0.3cm}
    >{\centering\arraybackslash}p{0.3cm}
    >{\centering\arraybackslash}p{0.3cm}
    >{\centering\arraybackslash}p{0.5cm}|
    >{\centering\arraybackslash}p{1.3cm}
    >{\centering\arraybackslash}p{1.3cm}
    >{\centering\arraybackslash}p{1.3cm}
    >{\centering\arraybackslash}p{1.3cm}
}

\toprule
\textbf{Paper} &
\textbf{Model(s)} &
\textbf{Funct.} &
\textbf{Task} &
\rotatebox{90}{\textbf{Security}} &
\rotatebox{90}{\textbf{Privacy}} &
\rotatebox{90}{\textbf{Reliability}} &
\rotatebox{90}{\textbf{Code Avail.}} &
\rotatebox{90}{\textbf{Accuracy \& Robustness}} &
\rotatebox{90}{\textbf{Risk Mgmt. System}} &
\rotatebox{90}{\textbf{Transparency \& Copy.}} &
\rotatebox{90}{\textbf{Vulnerability Eval.}} \\
\midrule

\cite{COSTA2024102880} & GPT-3.5-turbo & A & Route risk assessment for urban cyclists & \notaddressed & \notaddressed & \discussed & \codeyes & 1.50 $\pm$ 0.71 & 1.00$^{\dagger}$ & 1.00$^{\dagger}$ & 1.00$^{\dagger}$ \\
\cite{ZHANG202495} & GPT-3.5-turbo & A  & Traffic management & \implemented  & \discussed    & \implemented  & \codeyes & \textbf{3.50 $\pm$ 0.71} & 1.00$^{\dagger}$ & 1.00$^{\dagger}$ & 1.33 $\pm$ 0.58 \\
\cite{He_Nie_Ma_2025}  & LLaMA 3 8B, Mistral 8x7B & B   & Location representation   & \notaddressed & \notaddressed & \notaddressed & \codeyes & 2.00 $\pm$ 1.41 & 1.00$^{\dagger}$ & 1.00$^{\dagger}$ & 1.00$^{\dagger}$ \\
\cite{Muriuki2024AdvancedIT} & LSTM, GAN, RL    & A,D  & Traffic management & \discussed    & \discussed    & \discussed    & \codeno  & 1.00$^{\dagger}$ & 1.00$^{\dagger}$ & 1.00$^{\dagger}$ & 1.00$^{\dagger}$ \\
\cite{Kengne2025}  & BERT, TrajC-BERT variants   & D  & Trajectory prediction for carpooling \& taxi routing  & \notaddressed & \notaddressed & \discussed    & \codeyes & 2.00 $\pm$ 1.41 & 1.00$^{\dagger}$ & 1.00$^{\dagger}$ & 1.00$^{\dagger}$ \\
\cite{SinghAshrafRathore2025}& GPT-3   & A,D  & Traffic management & \notaddressed & \notaddressed & \notaddressed & \codeno  & 2.50 $\pm$ 0.71 & 1.00$^{\dagger}$ & 1.00$^{\dagger}$ & 1.33 $\pm$ 0.58 \\
\cite{sym17122083} & GPT-5 Thinking, GPT-4o, Gemini 2.5 Pro, DeepSeek V3 & C,D  & Traffic light optimization    & \notaddressed & \notaddressed & \notaddressed & \codeno  & 3.00 $\pm$ 2.83 & 1.00$^{\dagger}$ & 1.00$^{\dagger}$ & 1.00$^{\dagger}$ \\
\cite{ALSAHFI2025464}  & Gemma 2, Qwen 2.5, GPT-4o, Llama 3.2, Aya-Expanse   & A,D  & Traffic prediction & \notaddressed & \notaddressed & \notaddressed & \codeno  & 2.00 $\pm$ 1.41 & 1.00$^{\dagger}$ & 1.00$^{\dagger}$ & 1.00$^{\dagger}$ \\
\cite{GrigorevArtur2024}     & BERT-family models     & A  & Traffic incident severity classification & \notaddressed & \notaddressed & \notaddressed & \codeyes & 2.00 $\pm$ 1.41 & 1.00$^{\dagger}$ & 1.00$^{\dagger}$ & 1.00$^{\dagger}$ \\
\cite{Syum202511194151}      & GPT-3.5-turbo & A  & Real-time traffic insights     & \notaddressed & \notaddressed & \implemented  & \codeno  & 2.00 $\pm$ 1.41 & 1.00$^{\dagger}$ & 1.00$^{\dagger}$ & 1.00$^{\dagger}$ \\
\cite{s25216698}   & GPT-2   & D  & Traffic prediction & \discussed    & \discussed    & \discussed    & \codeno  & 2.00 $\pm$ 1.41 & 1.00$^{\dagger}$ & 1.00$^{\dagger}$ & 1.00$^{\dagger}$ \\
\cite{Liu202511005661} & GPT-2   & A,B  & Traffic prediction & \notaddressed & \notaddressed & \notaddressed & \codeyes & 2.00 $\pm$ 1.41 & 1.00$^{\dagger}$ & 1.00$^{\dagger}$ & 1.00$^{\dagger}$ \\
\cite{ChenYile2025}    & Llama 3.2-1B-Instruct & B     & & \notaddressed & \notaddressed & \implemented  & \codeyes & 2.50 $\pm$ 2.12 & 1.00$^{\dagger}$ & 1.00$^{\dagger}$ & 1.00$^{\dagger}$ \\
\cite{inffusYitongShang2025} & Meta-Llama    & A,D   & Electric vehicles charging demand prediction & \notaddressed & \notaddressed & \notaddressed & \codeno  & 2.50 $\pm$ 2.12 & 1.00$^{\dagger}$ & 1.00$^{\dagger}$ & 1.00$^{\dagger}$ \\
\cite{Calderon2025s25185688} & Qwen2.5-7B    & D   & Multimodal transportation simulation    & \notaddressed & \notaddressed & \discussed    & \codeyes & \textbf{3.50 $\pm$ 0.71} & 1.00$^{\dagger}$ & 1.00$^{\dagger}$ & 1.00$^{\dagger}$ \\
\cite{Kalantari2024}   & Llama 3 & A,D   & Fleet dispatching  & \implemented  & \notaddressed & \implemented  & \codeno  & 3.50 $\pm$ 2.12 & \textbf{3.00$^{\dagger}$} & 1.00$^{\dagger}$ & \textbf{3.67 $\pm$ 0.58} \\
\cite{Kalyuzhnaya2025} & GPT-4o, Mistral-8x22B, Llama 3.1   & D & Urban planning and smart-city management    & \discussed    & \discussed    & \notaddressed & \codeyes & 2.50 $\pm$ 2.12 & 1.00$^{\dagger}$ & 1.00$^{\dagger}$ & 1.00$^{\dagger}$ \\
\cite{FAN2026131620}   & GPT-2   & A,D  & Electric Vehicle Charging Forecasting   & \notaddressed & \implemented  & \notaddressed & \codeno  & 2.00 $\pm$ 1.41 & 1.00$^{\dagger}$ & 1.00$^{\dagger}$ & 1.00$^{\dagger}$ \\
\cite{ChenRuiqing2024} & GPT-3.5, GPT-4    & D   & Autonomous vehicle coordination   & \notaddressed & \notaddressed & \notaddressed & \codeno  & 2.00 $\pm$ 1.41 & 1.00$^{\dagger}$ & 1.00$^{\dagger}$ & 1.00$^{\dagger}$ \\
\cite{XU2025129562}    & GPT-2   & A,D  & Traffic flow prediction   & \notaddressed & \notaddressed & \notaddressed & \codeno  & 2.00 $\pm$ 1.41 & 1.00$^{\dagger}$ & 1.00$^{\dagger}$ & 1.00$^{\dagger}$ \\
\cite{LiZhonghang2024} & Vicuna-7B     & A,B   & Urban forecasting  & \notaddressed & \notaddressed & \implemented  & \codeno  & 2.50 $\pm$ 2.12 & 1.00$^{\dagger}$ & 1.00$^{\dagger}$ & 1.00$^{\dagger}$ \\
\cite{YuChenyangXie2024}     & LLaMA2-7B     & D  & Origin-Destination flow prediction & \notaddressed & \notaddressed & \notaddressed & \codeno  & 2.00 $\pm$ 1.41 & 1.00$^{\dagger}$ & 1.00$^{\dagger}$ & 1.00$^{\dagger}$ \\
\cite{CHENG2025113174} & BERT    & B,D  & Traffic flow prediction framework & \notaddressed & \notaddressed & \notaddressed & \codeno  & 2.00 $\pm$ 1.41 & 1.00$^{\dagger}$ & 1.00$^{\dagger}$ & 1.00$^{\dagger}$ \\
\cite{Li2025CausalII}  & Vicuna-7B, Vicuna-13B & D  & Traffic forecasting & \notaddressed & \notaddressed & \notaddressed & \codeyes & 2.00 $\pm$ 1.41 & 1.00$^{\dagger}$ & 1.00$^{\dagger}$ & 1.00$^{\dagger}$ \\
\cite{LouShangyu2025}  & GPT-4o  & A,B    & Urban forecasting  & \notaddressed & \notaddressed & \implemented  & \codeyes & \textbf{3.50 $\pm$ 0.71} & 1.00$^{\dagger}$ & \textbf{1.50 $\pm$ 0.71 }& 1.00$^{\dagger}$ \\
\cite{WangJingyuanWang2025}  & Unknown & B,D   & Point-of-interest recommendation & \notaddressed & \notaddressed & \notaddressed & \codeno  & 2.00 $\pm$ 1.41 & 1.00$^{\dagger}$ & 1.00$^{\dagger}$ & 1.00$^{\dagger}$ \\
\cite{ChenYong2025}    & GPT-4, Llama3-8B  & A,D    & Mobility perspective framework    & \discussed    & \discussed    & \discussed    & \codeno  & 2.50 $\pm$ 0.71 & 1.00$^{\dagger}$ & 1.00$^{\dagger}$ & 1.67 $\pm$ 1.15 \\
\cite{Bhandari2024}    & GPT-4-turbo, Gemini-Pro, Llama-2   & B    & Urban mobility assessment     & \notaddressed & \discussed    & \notaddressed & \codeyes & 2.00 $\pm$ 1.41 & 1.00$^{\dagger}$ & 1.00$^{\dagger}$ & 1.00$^{\dagger}$ \\
\cite{Masri2024LeveragingLL} & GPT-4o-mini   & D  & Traffic management & \notaddressed & \discussed    & \discussed    & \codeno  & 1.50 $\pm$ 0.71 & 1.00$^{\dagger}$ & 1.00$^{\dagger}$ & 1.00$^{\dagger}$ \\
\cite{Beneduce10981725}      & GPT-4o, GPT-4, GPT-3.5, Llama, Mistral families & D   & Zero-shot next-location prediction benchmark    & \notaddressed & \notaddressed & \notaddressed & \codeyes & \textbf{3.50 $\pm$ 0.71} & 2.00$^{\dagger}$ & 1.00$^{\dagger}$ & 1.00$^{\dagger}$ \\
\cite{Chen2025EnhancingTD}   & GPT-2   & D  & Taxi demand forecasting   & \notaddressed & \notaddressed & \notaddressed & \codeno  & 2.00 $\pm$ 1.41 & 1.00$^{\dagger}$ & 1.00$^{\dagger}$ & 1.00$^{\dagger}$ \\
\cite{Gan10112025} & BGE-Large, GLM4-9B, BERT     & B,D  & Traffic management regulations    & \notaddressed & \notaddressed & \notaddressed & \codeno  & 2.00 $\pm$ 1.41 & 1.00$^{\dagger}$ & 1.00$^{\dagger}$ & 1.00$^{\dagger}$ \\
\cite{HanQiuhan2025}   & Meta-Llama-3.1-8B & B,D   & POI prediction     & \notaddressed & \notaddressed & \notaddressed & \codeno  & 2.00 $\pm$ 1.41 & 1.00$^{\dagger}$ & 1.00$^{\dagger}$ & 1.00$^{\dagger}$ \\
\cite{HaoZhang11045306}      & GAN & A,D  & Traffic management & \discussed    & \notaddressed & \notaddressed & \codeno  & 3.00 $\pm$ 1.41 & 1.00$^{\dagger}$ & 1.00$^{\dagger}$ & 1.00$^{\dagger}$ \\
\cite{Madi202511104439}      & ST-LLM, GCN-GPT, GAT-GPT, Chronos     & D  & Traffic prediction & \notaddressed & \notaddressed & \notaddressed & \codeno  & 2.00 $\pm$ 1.41 & 1.00$^{\dagger}$ & 1.00$^{\dagger}$ & 1.00$^{\dagger}$ \\

\bottomrule
\end{tabular}
\begin{flushleft}
\scriptsize $^{\dagger}$ indicates that the standard deviation value is equal to zero ($\text{std} = 0.00$).\end{flushleft}
\caption{Review of Model(s) used, Functionality (Funct.), Task, Security, Privacy, Reliability, Code Availability, Accuracy \& Robustness, Risk Management System, Transparency \& Copyright, and Vulnerability Evaluation in surveyed papers.}
\label{table:survey}
\end{table*}